\documentclass{jaa}
\usepackage{natbib}
\usepackage{graphicx}
\usepackage{}

\newcommand{\hii}{\mbox{H{\sc ii}~}}

\begin{document}\sloppy

\title{H$\alpha$ emission line sources from VLT-MUSE in a low-metallicity star forming region - Dolidze 25}


\author{Mizna Ashraf\textsuperscript{1}, Jessy Jose\textsuperscript{1}, Gregory Herczeg\textsuperscript{2}, Min Fang\textsuperscript{3} }
\affilOne{\textsuperscript{1}Indian Institute of Science Education and Research, Tirupati. } \affilTwo{\textsuperscript{2}Kavli Institute of Astronomy and Astrophysics, China. } \affilThree{\textsuperscript{3}Purple Mountain Observatory, China.}

\twocolumn[{

\maketitle

\corres{mizna@students.iisertirupati.ac.in, jessyvjose1@gmail.com}



\vspace{5 mm}
\begin{abstract}

The process of accretion through circumstellar disks in young stellar objects is an integral part of star formation and the $H\alpha$ emission line is a prominent signature of accretion in low-mass stars. We present the detection and characterization of $H\alpha$ emission line sources in the central region of a distant, low-metallicity young stellar cluster - Dolidze 25 (at $\sim$ 4.5 kpc) - using medium-resolution optical spectra (4750-9350 \AA ) obtained with the Multi-Unit Spectroscopic Explorer (MUSE) at the VLT. We have identified 14 potential accreting sources within a rectangular region of (2$'$ x 1$'$) towards the center of the cluster based on the detection of strong and broad emissions in $H\alpha$ as well as the presence of other emission lines such as [OI] and $H\beta$. 
Based on their positions in both photometric color-magnitude and color-color diagrams, we have also confirmed that these objects belong to the pre-main sequence phase of star formation. Our results were compared with the disk and diskless members of the cluster previously identified by \citet{Guarcello2021} using near-IR colors, and all sources they had identified as disks were confirmed to be accreting based on the spectroscopic characteristics. 
\end{abstract}


}]


\doinum{12.3456/s78910-011-012-3}
\artcitid{\#\#\#\#}
\volnum{000}
\year{2022}
\pgrange{1--}
\setcounter{page}{1}
\lp{1}

\section{Introduction}

Protoplanetary disk evolution plays a pivotal role in the theory of star and planet formation \citep[see reviews by][]{Andrews2020, Manara2022}. Disks form due to the gravitational collapse of pre-stellar cloud cores and then disperse as the host star progresses toward the main sequence phase. Throughout their existence, protoplanetary disks go through various physical processes that cause their gas and dust composition and their structural properties to change.\par 

During the early phase of the pre-main-sequence (PMS), young stars undergo active accretion of matter from the surrounding optically thick, gas-rich disk \citep{1998Hartmann}. This phase is marked by various emission lines and UV/optical continuum excess, resulting from the accretion of high angular momentum material from the disk onto the central star. The near-IR and mid-IR regimes also exhibit excess continuum dust emission from the heated inner disk. \par

One of the most common signatures of accretion in low-mass stars is a broad, asymmetric, and typically strong $H\alpha$ emission line. The strength of this line can distinguish between classical T Tauri stars or accretors and weak-line T Tauri stars or non-accretors \citep{Muzerolle2003}. In the context of magnetospheric accretion, the $H\alpha$ line is created due to gas falling onto the star from the inner disk \citep{Hartmann2016}. The strength of the $H\alpha$ emission line decreases with increasing stellar age, indicating a gradual decline in accretion activity during the PMS phase. Eventually, as stars evolve into weak-lined T Tauri stars or non-accretors, their $H\alpha$ emission weakens and arises primarily from chromospheric activity. The $H\alpha$ profiles of accretors tend to be broader due to the high velocity and opacity of the accreting gas and may exhibit asymmetries resulting from inclination effects or absorption by a wind component \citep{Kurosawa2008}. The presence or absence of accretion signatures offers valuable insight into the inner disk evolution of young stellar objects. \par

Numerous studies show that a variety of factors may have an impact on how the disk evolves, which include stellar and disk mass, metallicity, and the local environment \citep{Hartmann2016, Guarcello2021, Fang2009, Fang2012}. The initial metallicity of disks is thought to play a significant role in star formation and disk evolution by influencing the relative content of dust \citep{Yasui2009, 2016aYasui, 2016bYasui, 2021Yasui}. Observations on the accretion process in star-forming regions of low metallicity, both within our galaxy and in extragalactic systems, have revealed that on average, low-metallicity stars exhibit higher accretion rates than their solar metallicity counterparts \citep{2012Spezzi, 2013Demarchi}. However, a conflicting result was obtained by \citet{Kalari2015}, who found no substantial difference in the inferred mass accretion rates between low-metallicity $(1/6\,{Z}_{\odot })$ stars in the Dolidze 25 open cluster and solar metallicity pre-main sequence stars of equivalent mass. A further in-depth examination is necessary to determine the source of this discrepancy.\par

The young open cluster Dolidze 25 (Do 25) is one of the best-known rare cases of galactic low-metallicity environments, with average abundances of $\sim$ 0.5 to 0.7 dex below solar metallicity \citep{Lennon1990, Fitzsimmons1992, Negueruela2015}. Located at a distance of $ \sim 4.5$ kpc towards the galactic anticentre and associated with the \hii region Sh2-284 \citep{Puga2007}, Do 25 has an estimated age of 1 to 2 Myr \citep{Guarcello2021}. Do 25 is  thus an ideal laboratory to study the role of metallicity in star formation.\par

The main objective of this research is to gain insights into the impact of metallicity on the evolution of protoplanetary disk in star formation. In this paper, we present our findings from the search for $H\alpha$ emission line sources in the core of the Dolidze 25 cluster using VLT-MUSE IFU observations. Our analysis is supported by the existing list of disk and diskless members of the cluster previously identified by \citet{Guarcello2021}. The paper is organized as follows: Section 2 provides an overview of the observations, Section 3 focuses on the detection of $H\alpha$ emission line sources in the cluster, Section 4 characterizes the identified members, and finally, Section 5 summarizes the results. A more comprehensive analysis of the impact of metallicity on protoplanetary disk evolution will be presented in a future publication.

\begin{figure}
    \centering
    \includegraphics[scale=0.25]{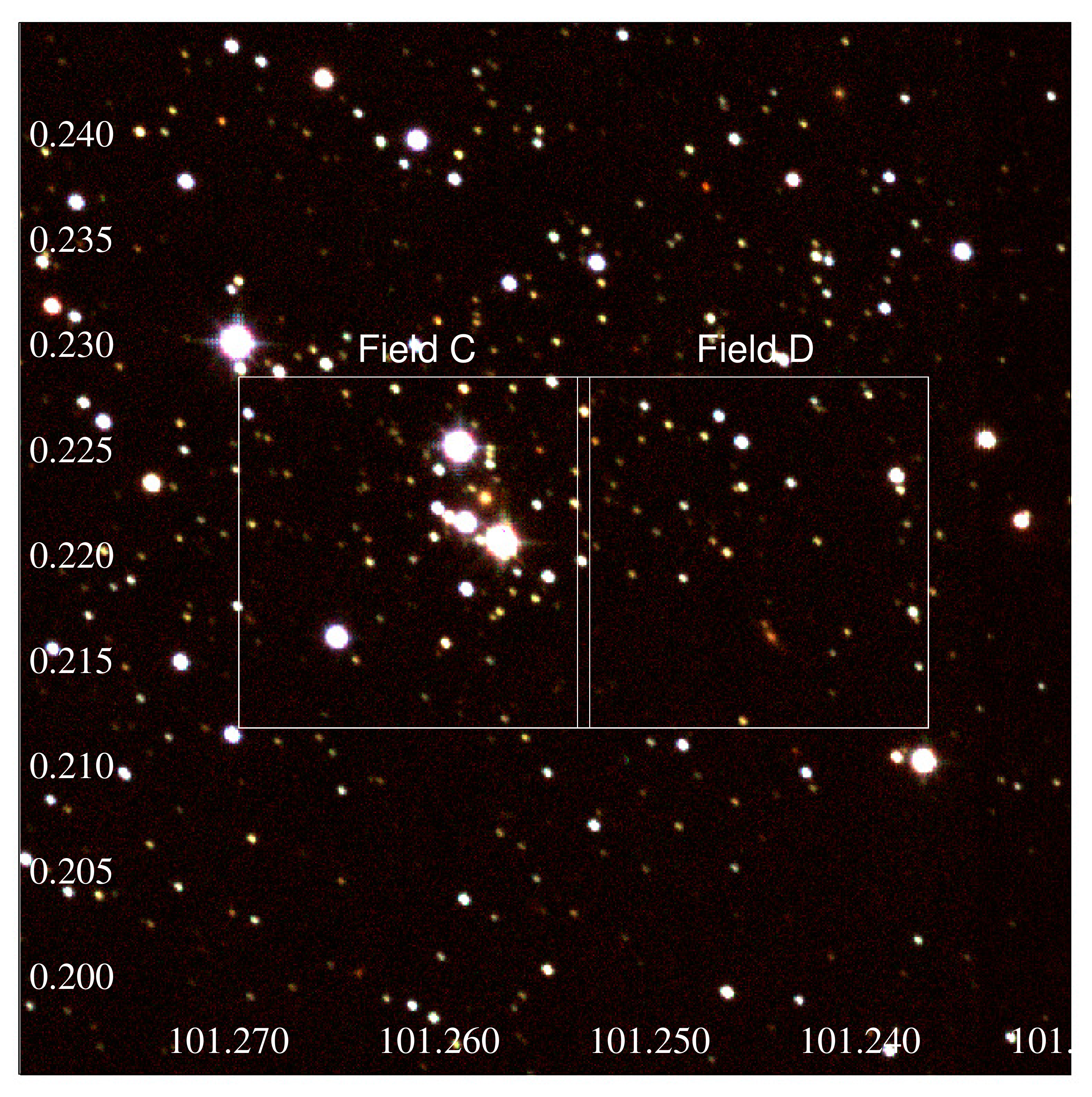}
    \caption{RGB image of the central area of Dolidze 25 constructed from the UKIDSS JHK bands. The MUSE field of view of two pointings (C and D) are represented by white boxes, which together cover an area of 2$'$ x 1$'$ around the center of the cluster.}
\end{figure}

\section{Observations from VLT-MUSE}
The raw data for Dolidze 25 observed during January 2017 using the  VLT-MUSE integral field unit (IFU) (PI: G. Herczeg; Program ID: 098.C-0435) was obtained from the ESO science archive facility. The observations were taken in the nominal wavelength range (4750 - 9350 \AA) using the instrument's wide-field mode (1$'$ x 1$'$ per pointing). The spectral resolution of MUSE ranges from 1750 at 4750 \AA~ to 3750 at 9350 \AA, at $H\alpha$ (6563 \AA) the resolving power is $\sim 2500$. Two pointings were obtained towards the core of the cluster centered at (06:45:02.68, +00:13:11.4 ; Field C) and (06:44:58.82, +00:13:11.4; Field D) (see Figure 1). The seeing during the night was $\sim$ 0.7 arcseconds.  \par

\begin{figure*}
    \centering
    \includegraphics[scale=0.12]{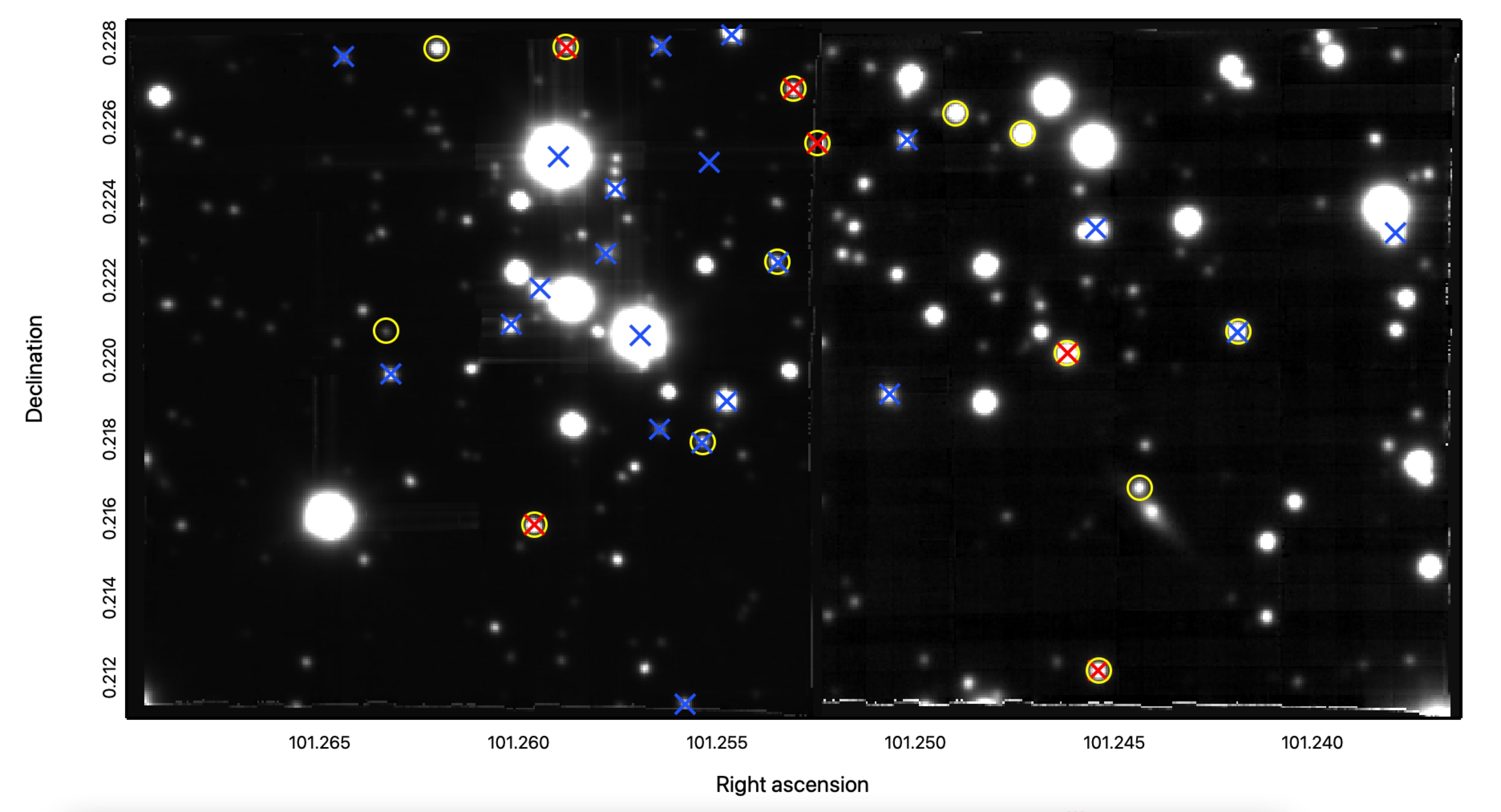}
    \caption{Integrated and mosaiced $i$-band image obtained from the data cube of  MUSE-IFU observations. The yellow circles represent the $H\alpha$ emission line sources from MUSE spectral analysis. Red crosses are disk sources, and blue crosses are diskless sources identified based on their NIR excess/non-excess characteristics by  \citet{Guarcello2021}.}
    \label{museimage}
\end{figure*}

This data was reduced using the MUSE EsoRex pipeline recipes (2.8.4) \citep{Weilbacher2014} to obtain the wavelength and flux-calibrated data cubes. However,  upon inspection, the cubes were found to have poorly subtracted sky with negative fluxes because of heavy contamination present due to the skylines. The sky subtraction method present in the pipeline is not ideal for nebular regions since it considers some of the nebular emissions as skylines \citep{Anna2020}. This resulted in negative fluxes in the reduced data cube. Therefore, we re-reduced the raw data utilizing the modified sky subtraction method by MusePack \citep{Peter2019}, a wrapper for the standard reduction pipeline. Figure \ref{museimage} shows the integrated and mosaiced i-band image constructed using the MUSE datacubes from two pointings.  \par

Since Do 25 is associated with the \hii region Sh2-284, emission from the nebular background needs to be completely removed to ensure that the observed lines are solely from the star. We eliminated the nebular emission from the \hii region by subtracting the local background 4$\times$FWHM away from each stellar source. We inspected the individual spectra using the [N II] $( \lambda \sim 6549, 6583)$ lines as a gauge for the quality of the background removal. These two lines are purely from the nebula and should be completely absent in the stellar spectra \citep{Peter2019}. Thus we ensure that the emission lines associated with the stellar spectra are devoid of any background contribution. Details on sky subtraction, wavelength and flux  calibration and the calibration accuracy will be presented elsewhere. \par
The spectra of 290 sources were extracted from the total field of view. For the 183 spectra with the signal-to-noise ratio greater than 5, we shifted the wavelengths adopting radial velocity of the cluster to be 71 km/s based on  \citet{Wu2009}. We use these spectra to  identify sources with $H\alpha$ emission. In addition, we utilised archival optical and near-IR photometry from Pan-STARRS \citep{Panstarrs2016}, IPHAS DR2 \citep{IPHAS2014}, and UKIDSS DR10 \citep{UKIDSS2007} for  further analysis.

\begin{figure*}
    \centering
    \includegraphics[scale=0.021]{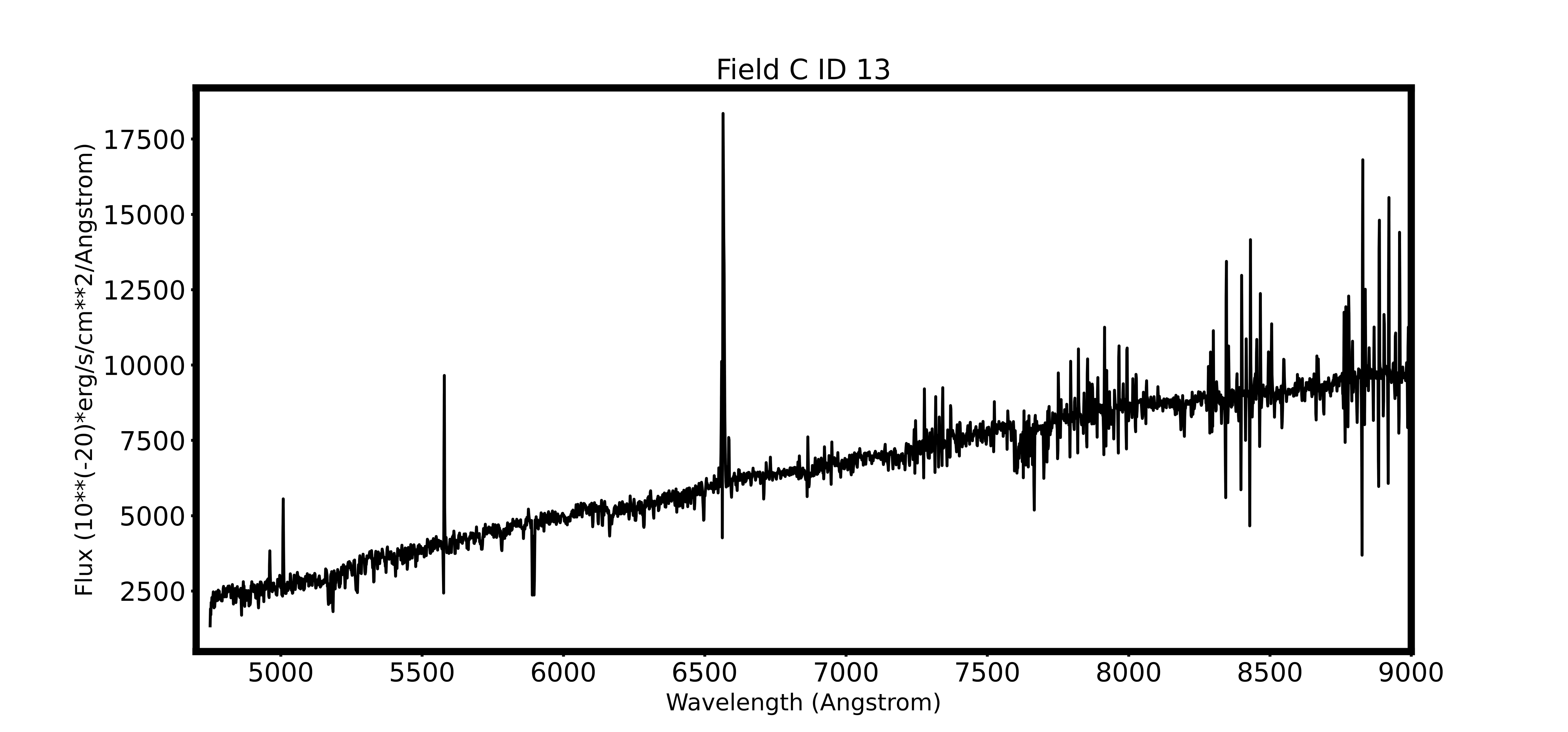}
    \includegraphics[scale=0.021]{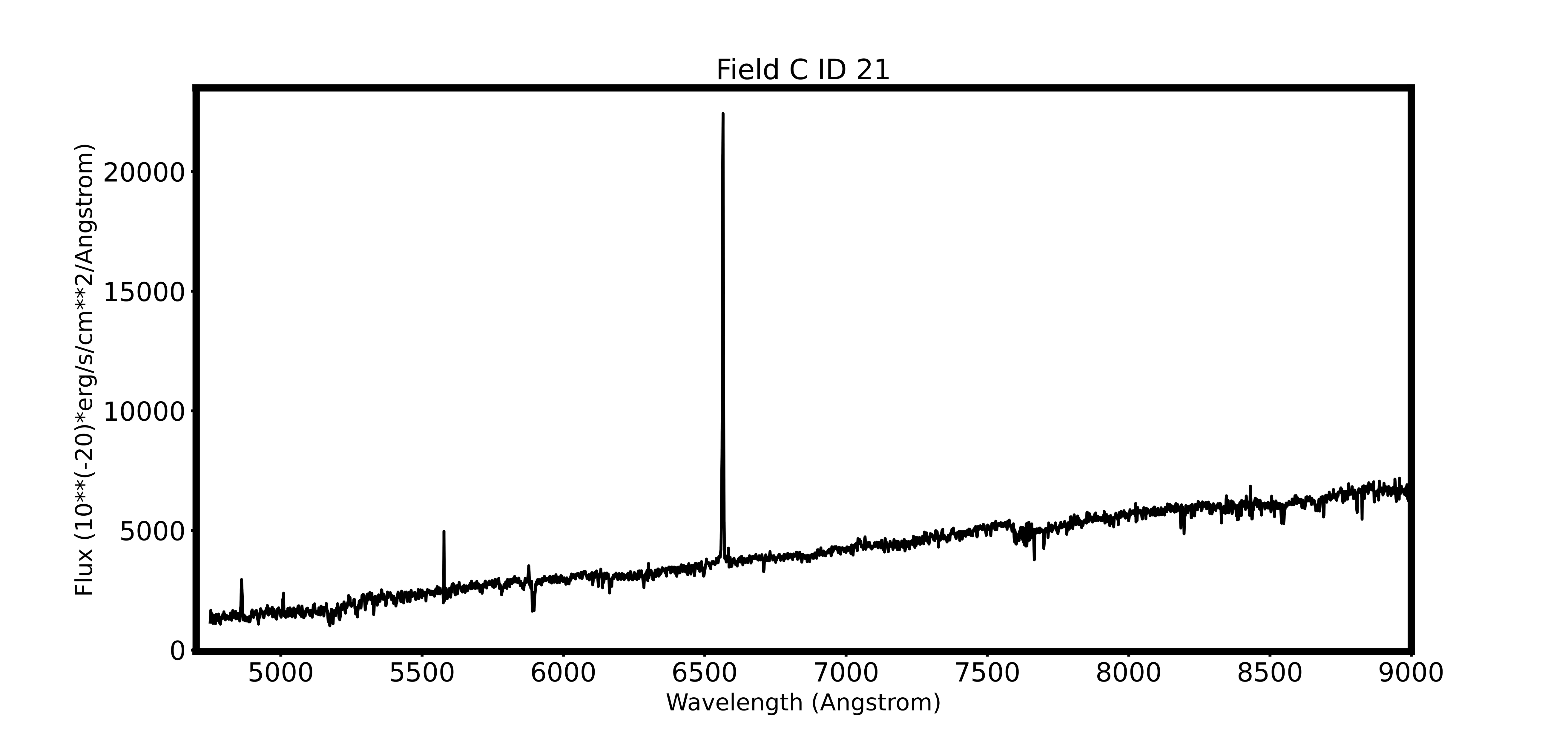}
    \includegraphics[scale=0.021]{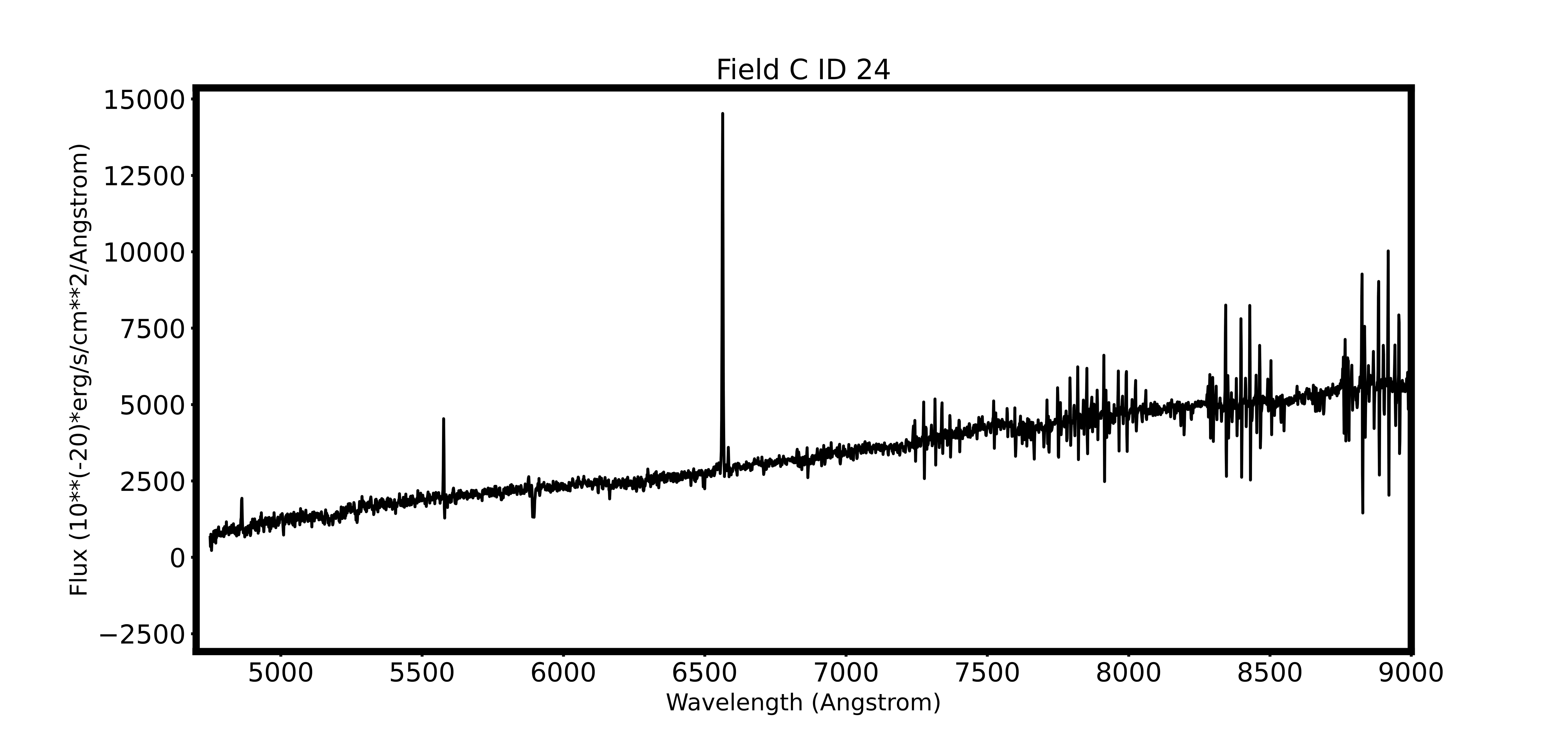}
    \includegraphics[scale=0.021]{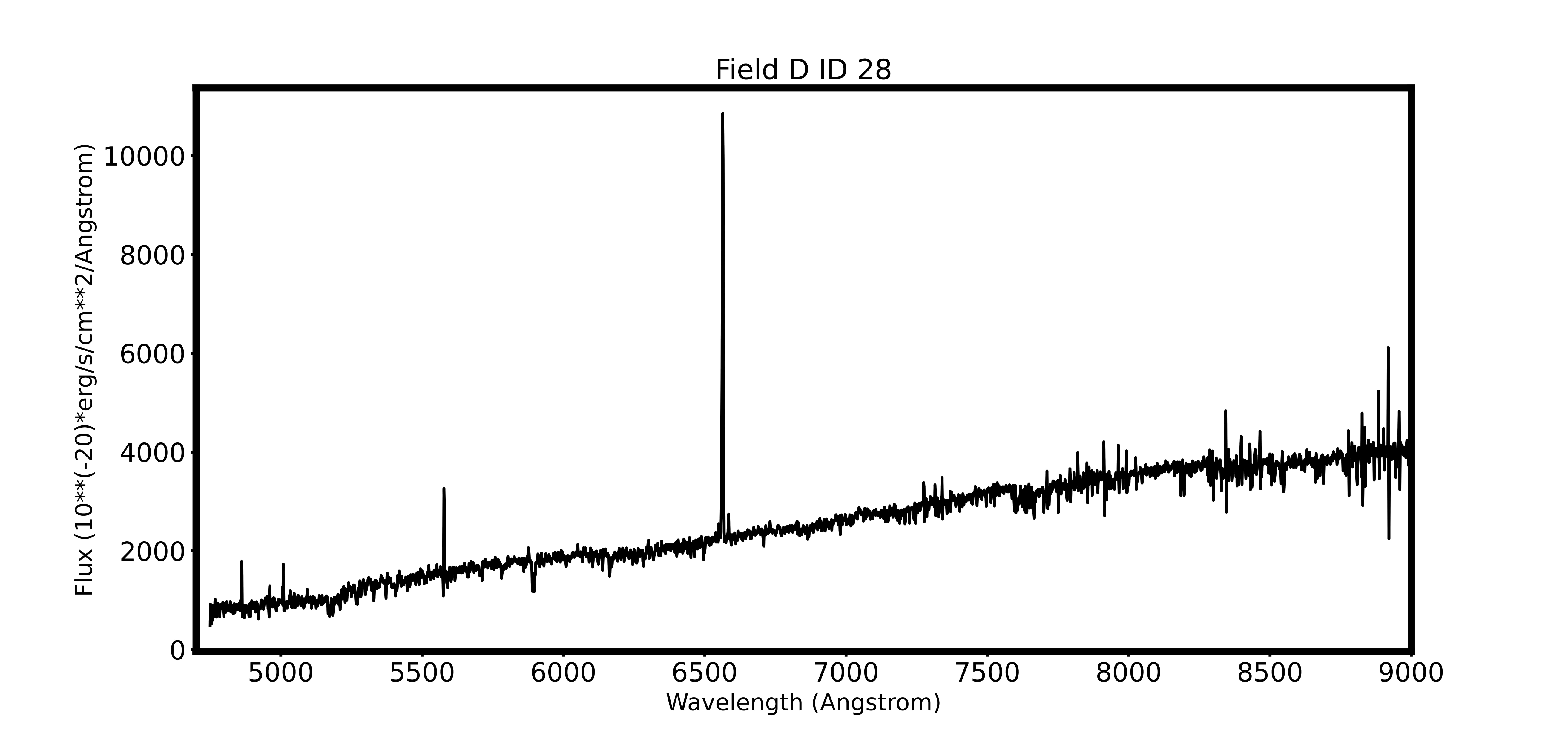}
    \includegraphics[scale=0.021]{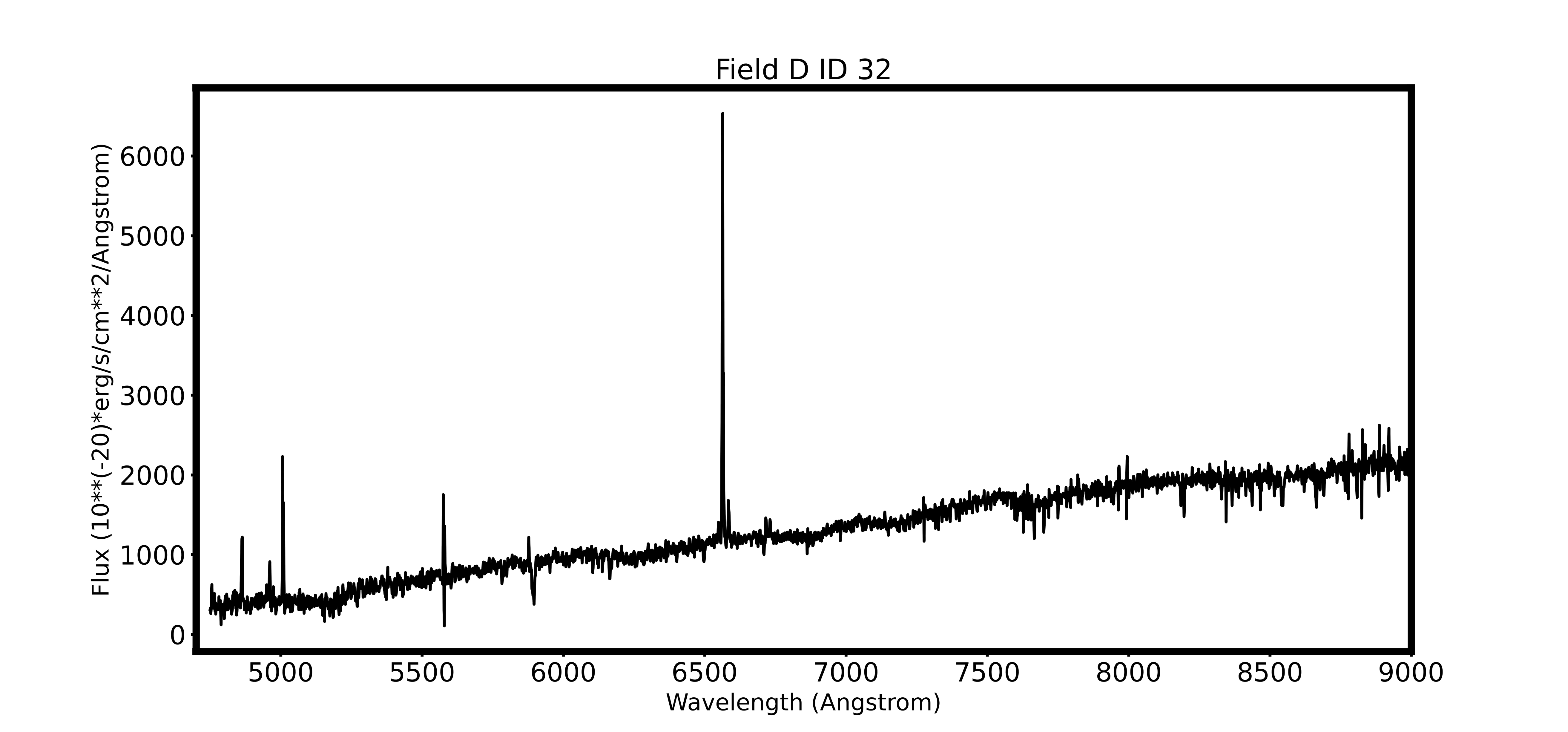}
    \includegraphics[scale=0.021]{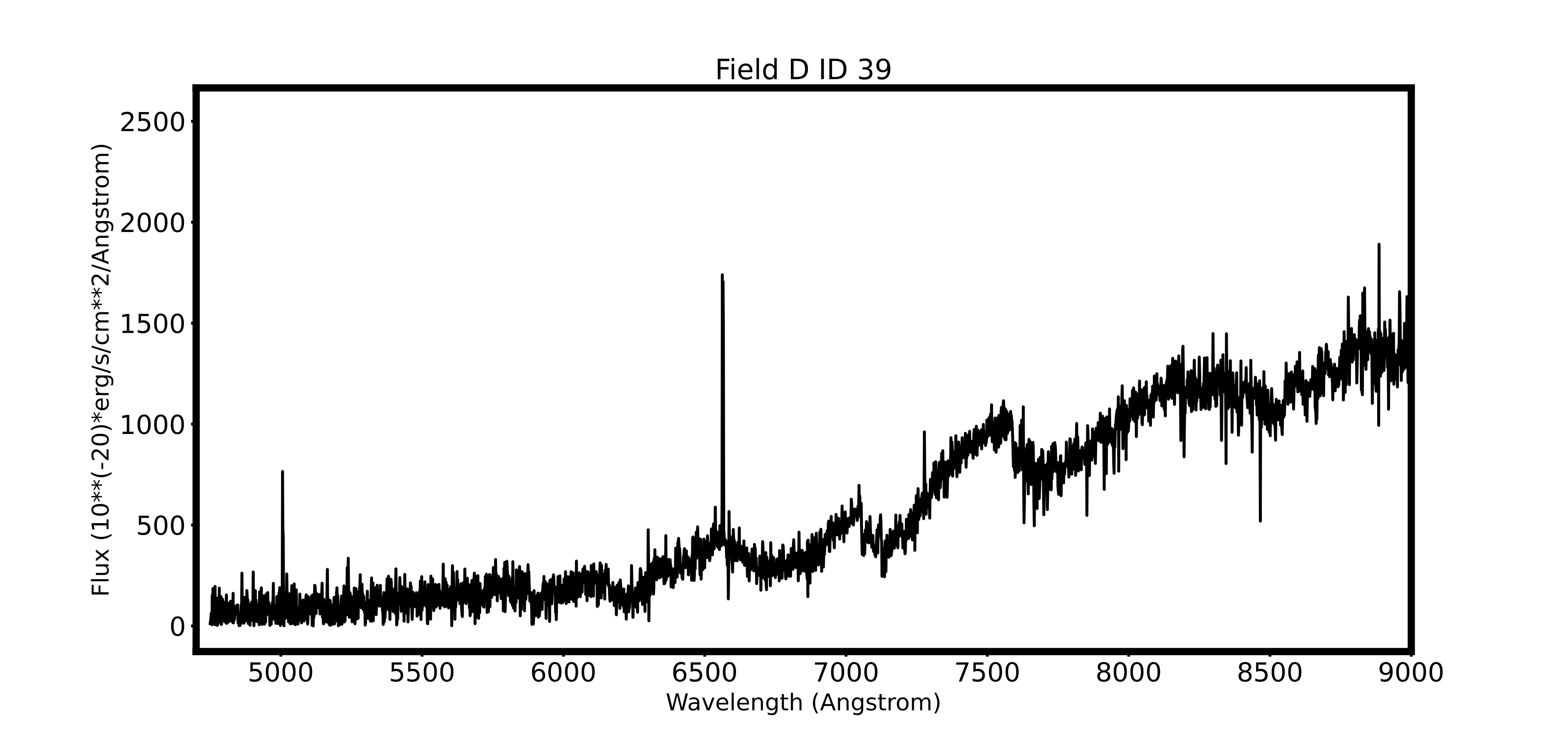}
    \caption{Spectra of accreting sources which are confirmed as Disk bearing members of the cluster based on NIR excess emission.}
    \label{spectra}
\end{figure*}

\begin{table*}[htb]
\centering
\tabularfont
\caption{Details of sources with emission in $H\alpha$ identified using MUSE-IFU observations }
\label{table} 

\begin{tabular}{lccccccc}

\topline

RA & Dec & J (mag) & H (mag) & K (mag) & EW $H \alpha (\AA) $  & 10\% width (km/s) \\
\midline
 101.24456 & 0.21642 & 18.197 $\pm$0.048 & 17.267 $\pm$ 0.038 & 16.588 $\pm$ 0.047 & -52 $\pm$ 3.0 & 495  \\
 101.26342 & 0.22046 & 18.200 $\pm$0.049 & 17.253 $\pm$ 0.037 & 16.614 $\pm$ 0.048 & -43 $\pm$ 4.0 & 374 \\
 101.25261 & 0.22512 & 18.080 $\pm$0.044 & 16.917 $\pm$ 0.027 & 16.289 $\pm$ 0.035 & -38 $\pm$ 3.0 & 365 \\
 101.25312 & 0.22653 & 16.219 $\pm$0.008 & 15.282 $\pm$ 0.006 & 14.667 $\pm$ 0.008 & -27 $\pm$ 0.4 & 450\\
 101.24560 & 0.21182 & 17.270 $\pm$0.021 & 16.308 $\pm$ 0.016 & 15.781 $\pm$ 0.023 & -26 $\pm$ 0.5 & 270\\
 101.24636 & 0.21983 & 16.737 $\pm$0.013 & 15.801 $\pm$ 0.010 & 15.258 $\pm$ 0.014 & -19 $\pm$ 0.4 & 422\\
 101.24206 & 0.22036 & 17.340 $\pm$0.022 & 16.460 $\pm$ 0.018 & 16.163 $\pm$ 0.032 & -19 $\pm$ 0.6 & 384\\
 101.25885 & 0.22761 & 16.310 $\pm$0.009 & 15.307 $\pm$ 0.006 & 14.611 $\pm$ 0.008 & -18 $\pm$ 0.5 & 391\\
 101.24913 & 0.22590 & 17.162 $\pm$0.019 & 16.262 $\pm$ 0.015 & 15.913 $\pm$ 0.025 & -13 $\pm$ 0.5 & 532\\
 101.25972 & 0.21554 & 15.980 $\pm$0.007 & 15.179 $\pm$ 0.005 & 14.785 $\pm$ 0.009 & -13 $\pm$ 0.5 & 462\\
 101.24744 & 0.22536 & 16.930 $\pm$0.015 & 16.139 $\pm$ 0.013 & 15.885 $\pm$ 0.024 & -12 $\pm$ 0.5 & 536 \\
 101.25545 & 0.21764 & 16.929 $\pm$0.015 & 16.059 $\pm$ 0.012 & 15.719 $\pm$ 0.021 & -10  $\pm$ 0.5 & 376\\
 101.25359 & 0.22218 & 16.604 $\pm$0.012 & 15.753 $\pm$ 0.009 & 15.394 $\pm$ 0.016 & -4.0 $\pm$ 0.5 & 320\\
 101.26211 & 0.22759 & 16.454 $\pm$0.010 & 15.675 $\pm$ 0.009 & 15.392 $\pm$ 0.016 & -3.0 $\pm$ 0.5 & 431\\

\hline
\end{tabular}
\end{table*}

\section{Emission line sources and young candidate members in the cluster}

In this paper, we identify sources with significant emission in $H\alpha$ towards the center of Do 25. We compare our results with the disk and diskless sources identified within the MUSE field of view from literature based on their NIR excess/non-excess properties. Below we detail the methods. 

\subsection{H$\alpha$ emission line sources within Do 25}
All spectra in our dataset that had a signal-to-noise ratio of 5 or greater were inspected for the presence of the H$\alpha$ emission line feature. Any spectrum that exhibited a significant emission line at $\lambda \sim 6563 \AA$ was categorized as a candidate accretor. The equivalent width(EW) of the H$\alpha$ line was then determined for all candidate accretors in our sample using PHEW \citep{PHEW2022}. This utility uses PySpecKit \citep{pyspeckit} to automate the Equivalent width calculation to fit a Voigt profile to the H$\alpha$ line. By conducting 1000 Monte Carlo iterations with Gaussian noise introduced to the flux spectrum, PHEW determines the mean and standard deviation of the resulting 1000 EWs, which we adopt as the measured EW and its 1$\sigma$ uncertainty.\par

The measured EW of H$\alpha$, and consequently the threshold for classifying an object as an accretor, depends on the spectral type. \citet{White2003} demonstrated that the full width at 10\% of the H$\alpha$ emission profile peak provides a more practical and potentially more precise indicator of accretion than the equivalent width of H$\alpha$ or optical veiling. Stars with 10\% width $>$ 270 km/s are regarded as accreting sources (classical T Tauri stars), regardless of their spectral type. To determine the 10\% width of the H$\alpha$ line from the MUSE spectrum, we adopted a technique similar to the one outlined by \citet{Fedele2010} for obtaining 10\% widths from low-resolution VIMOS spectroscopy. This method involves drawing a horizontal line at the evaluation height, which corresponds to 10\% of the line peak, and extending it on both sides until it intersected with the signal. We then estimated the wavelengths corresponding to the intersection points using linear interpolation and calculated the difference between them as the 10\% width. We measured the 10\% width of $H\alpha$ line for all sources with an equivalent width greater than $3 \AA$ using this method. In low signal-to-noise spectrum, the 10\% level might be confused with the continuum fluctuations. To avoid this, all the measured widths were visually inspected.\par

Possible accretors are selected based on the shape of their line profiles, specifically whether the H$\alpha$ emission line is strong (EW $>$ 3 $\AA$) and asymmetric with 10\% width $>$ 270 km/s, as well as the existence of emission lines such as [OI] and H$\beta$, which are often associated with accretion disk-driven outflows \citep{Hartigan1995}. We thus identify 14 sources within the field of view of Do 25 falling under this criteria.  The $H\alpha$ emission line sources thus identified are marked in  Figure \ref{museimage}, with properties listed in Table \ref{table}. Figure \ref{spectra} shows sample spectra of some of the $H\alpha$ emission line sources identified within the cluster. 


\begin{figure*}[!ht]
    \centering
    \includegraphics[scale=0.052]{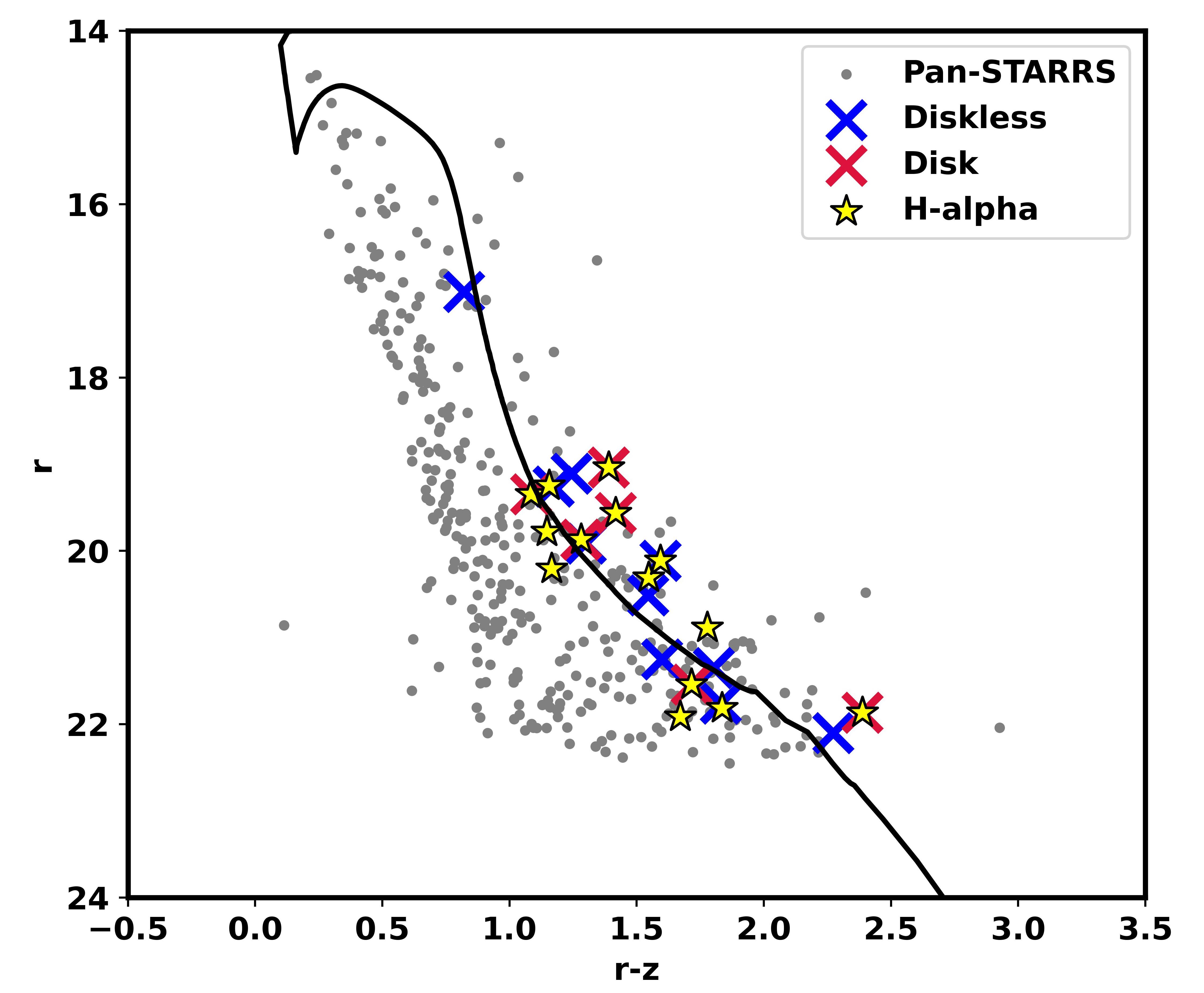}
    \hspace{6 mm}
    \includegraphics[scale=0.052]{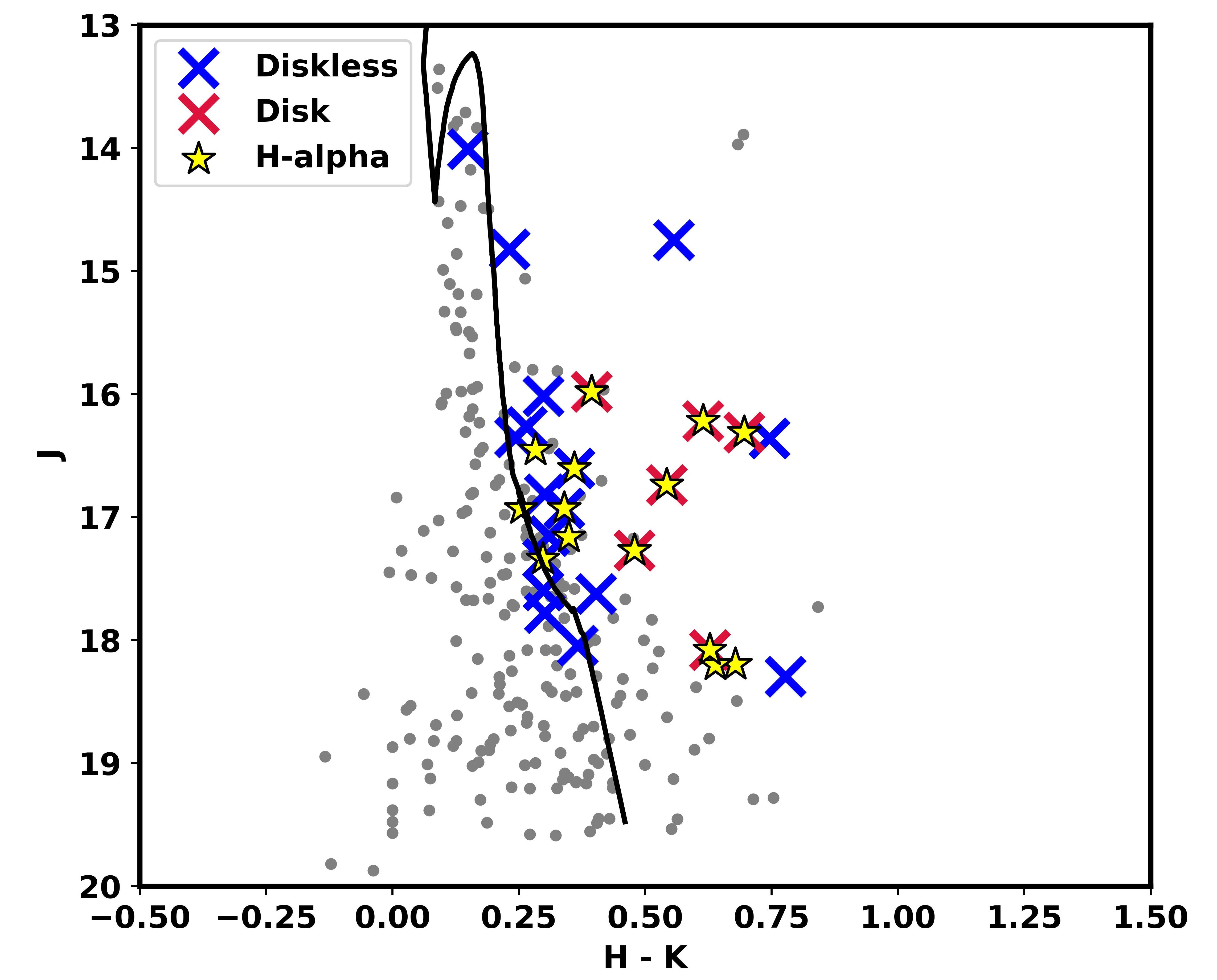}
    \caption {Left: $r-z$ vz $r$ optical Color-Magnitude Diagram using pan-STARRS photometry. Right: Infrared Color–magnitude diagram using UKIDSS JHK photometry. The red crosses and blue crosses indicate disk members and diskless members, identified by \citet{Guarcello2021}, and yellow star marks are the $H\alpha$ emission line sources within the MUSE FoV.  Black solid lines represent the 1 Myr PARSEC  isochrones \citep{Parsec2012} at a metallicity of Z = 0.003 corrected for an $A_V$ of 2 mag and distance of 4.5 kpc. }
    \label{CMD}
\end{figure*}

\subsection{Disk bearing and diskless sources}

We compare the list of $H\alpha$ emission line sources identified from MUSE spectra with the disk bearing and diskless sources from \citet{Guarcello2021}. They used infrared and optical photometry from UKIDSS/2MASS, Spitzer-IRAC, PanSTARRS, IPHAS/VPHAS along with X-ray data from $\it{Chandra}$ space telescope to identify the candidate disk and diskless members of the cluster. They provide a catalog of the selected disk-bearing population in a 1$^\circ$ circular region centered on Dolidze 25 from the criteria based on infrared colors and the disk-less population within a smaller central region based on X-ray sources with optical and IR counterpart (see Figure \ref{CCD}). They derived the disk fraction from the smaller central region of the cluster to be $\sim 34 \pm 4 \% $. The photometric sensitivity of the above surveys largely limits their study. \par

\begin{figure*}[h!]
    \includegraphics[scale=0.36]{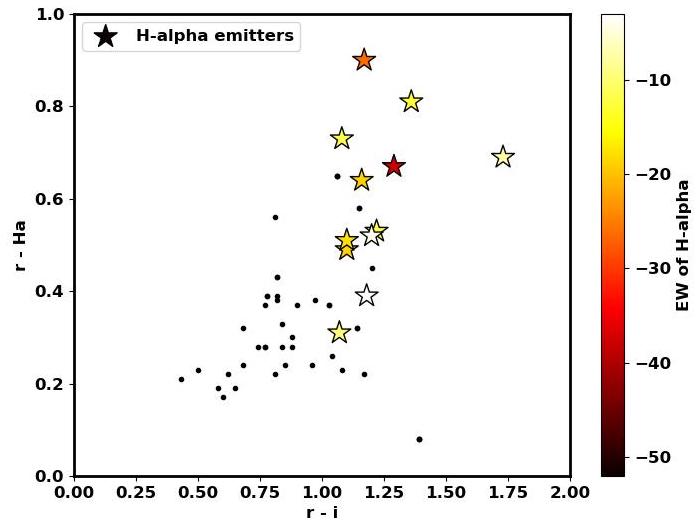}
    \hspace{2 mm}
    \includegraphics[scale=0.36]{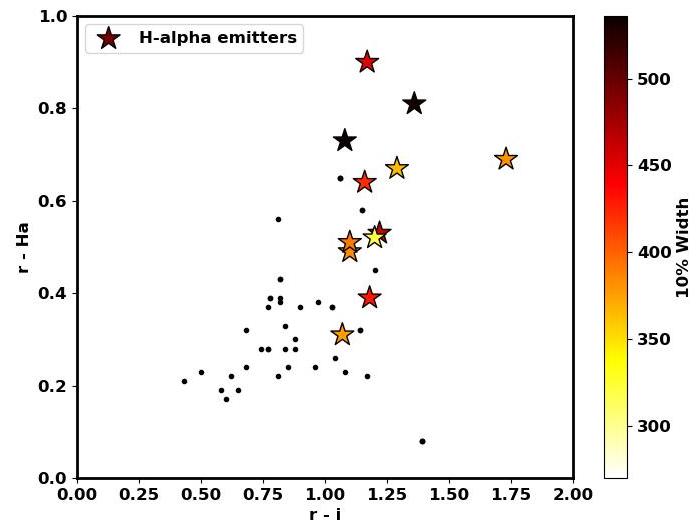}
    \caption {($r-H\alpha$) vs ($r-i$) color-color diagram using IPHAS photometry for sources within the MUSE field of view.  $H\alpha$ emission line sources identified from MUSE spectra are marked as stars. The color of the emission line sources represents their $H\alpha$ EW (left), and 10\% width (right), and the respective color bars are also shown.  }
    \label{halpha}
\end{figure*}

All the disk and diskless sources reported in \citet{Guarcello2021} have been retrieved in our MUSE data within its (2$'$ x 1$'$) field of view. There are six disks and 21 diskless sources confirmed as members of the cluster by \citet{Guarcello2021} within the FOV of MUSE. We search for their counterparts in the MUSE data, and  all six disk sources are included in the list of $H\alpha$ emission line sources. In addition, 3 of the 21 diskless members also show $H\alpha$ emission  in MUSE spectra. These sources  have 10\% width $>$ 300 km/s, and hence we include them in the list of possible accretors. Complementary to the spectroscopic characteristics, they exhibit a significant excess in IPHAS/VPHAS-based r-$H\alpha$ colors (see figure \ref{halpha}), which provide additional evidence that they are candidate accretors. \par

Additionally, we found five more sources in $H\alpha$ emission, which are not recognized as cluster members by \citet{Guarcello2021}. The membership of these five sources was likely not established as they were not detected in the IR/X-ray catalog. Since the MUSE observations are deeper than IRAC photometry, we are identifying more accretors though they are not listed in the disk sources in \citet{Guarcello2021}. The spatial locations of these sources are marked in Figure \ref{museimage}.

\section{Discussion}
\begin{figure*}[ht]
    \centering
    \includegraphics[scale=0.6]{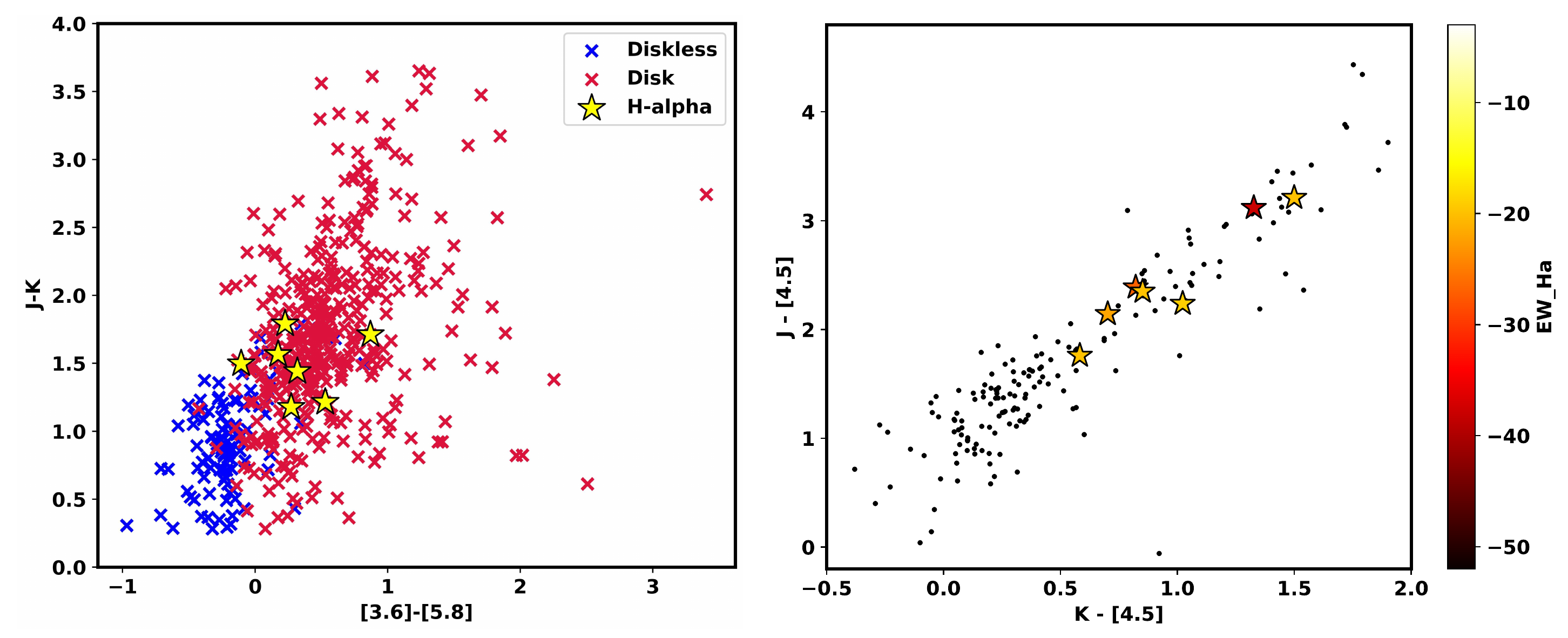}
    \caption {Left: UKIDSS J-K vs IRAC [3.6]-[5.8] color-color diagram. The red and blue crosses are the disk and diskless members identified by \citet{Guarcello2021} within an area of  $\sim$ 1$^\circ$ around Do 25; they are plotted to easily identify the  segregation of two classes. Right: J-4.5 vs K-4.5 color-color diagram of sources within the central region of Do 25. The yellow stars indicate the accretors identified using MUSE spectra color coded based on the their EW. Only 7 out of 14 accretors have detections in the IRAC [3.6], [4.5] and [5.8] bands.}
    \label{CCD}
\end{figure*}

The selection of accretors was made by taking into account all of the good SNR spectra that were retrieved from the cube, but no membership analysis was carried out. We obtained the archival photometry from Pan-STARRS, which has the deepest data available in the optical wavelength range; IPHAS, for the magnitudes measured with $H\alpha$ filter; and UKIDSS, which has the deepest infrared data available for the region. We inspected the location of the accretors on the  color-magnitude and color-color diagrams along with the isochrones matching with the age  ($\sim$ 1 Myr), metallicity ($\sim$ 0.003) which is corrected for the extinction ($A_V$ $\sim$ 2mag) and  distance (4.5kpc) \citep{Guarcello2021} of the cluster. \par

The color-magnitude diagram (CMD) using Pan-STARRS optical photometry as well as the infrared CMD using UKIDSS photometry is given in Figure \ref{CMD}. Both of the diagrams show that especially in the NIR CMD, the diskless sources are mostly less scattered and lined up along a sequence close to the isochrone, whereas the identified accretors are mostly scattered towards  the right of the pre-main sequence branch, which indicates that they belong to the pre-main sequence class of stars. Since all the accreting sources are positioned along the pre-main sequence branch in both CMDs along with the disk and diskless cluster members identified in the literature, the $H\alpha$ emission line sources can be considered as candidate members of the cluster. Further spectroscopic analysis is essential to confirm their membership, which is beyond the scope of this paper. 

Figure \ref{halpha} shows the $r-i$ versus $r-H\alpha$ color-color diagram for all  the sources in the field of view of MUSE for which IPHAS photometry is available. H$\alpha$ equivalent width, as well as 10\% width measured from MUSE spectra, are represented as color gradient in the figure. The general trend reveals that the spectroscopic equivalent widths and 10\% widths are proportional to the photometric color excess in IPHAS $r-H\alpha$ on the Y-axis of Figure \ref{halpha}. It is to be noted that the photometric and spectroscopic measurements are from two separate epochs, hence it is assumed that fluctuations in the trend are likely to be caused by variability in accretion \citep[see review by][]{Fischer2022}.

To confirm the spectroscopic classification of stars with disks from those without disks, in figure \ref{CCD}, we plot the Spitzer/IRAC mid-infrared photometric colours against J-K colours from UKIDSS, a good proxy for the stellar photosphere as well as excess emission from the disk. To better visualize the segregation between the two classes of sources, we also plot the disk-bearing members within a 1$^\circ$ circular region centered on Dolidze25 and the disk-less population within a smaller central region published in the \citet{Guarcello2021} catalog. The color-color diagrams (Figure \ref{CCD}) differentiate the two classes of objects. Diskless stars do not exhibit an excess of color in the IRAC channels and are therefore located near the origin, whereas stars with disks are clustered in an area with an excess of color. Only 7 of the 14 identified accreting sources have been detected in the spitzer 3.6, and 4.5, 5.8 $\mu m$ bands, while only 6 have been detected in the 8 $\mu m$ band. These sources have been identified as having infrared excess, and their positions in the color-color diagram correspond to those of disk sources. \par

Several studies have shown that on average, the emission-line strength in pre-main-sequence stars decreases with age and decreases significantly in a short period of time \citep{2006Manoj}. Additionally, stars with low emission-line activity tend to have smaller IR excess, while those with strong emission-line activity exhibit larger IR excess \citep{1990Cabrit}. In our analysis, we find that the sources with low equivalent widths are not detected in the IRAC bands, suggesting a lower level of IR excess, which aligns with previous studies. \par
Figure \ref{CCD} (right panel) presents K - [4.5] and J - [4.5] color-color diagram. The sources identified as H$\alpha$ emitters in MUSE spectra are shown as star symbols color coded based on the their H$\alpha$ EWs. Of the 14 H$\alpha$ emitters, only 7 have detections in the IRAC [4.5] band, and these are the sources with H$\alpha$ equivalent widths greater than $\sim -20$ in the MUSE spectra. Considering the correlation between the NIR excess and the strength of the H$\alpha$ emission, the sources with lower EWs may not be brighter at NIR wavelengths. Furthermore, the emission lines are considered to be a better indicator of accretion activity than the infrared excess, as the latter can be produced even by passive disks through the presence of hot ($T \sim 1000 K$) dust close to the star \citep{2004Muzerolle}. Therefore, we cannot affirm that the sources not detected by Spitzer are WTTS. It is also noteworthy that the emission line activity can exhibit substantial fluctuations in a short amount of time due to accretion variability. Although all 14 sources are considered to be CTTS based on their spectroscopic characteristics, we cannot rule out the possibility of some of them being WTTS, given their smaller EW values.

\section{Conclusion and Future Prospects}

Low-metallicity protoplanetary discs are hypothesized to develop on different timescales compared to solar-metallicity ones. Despite its relevance, the conclusions on the impact of metallicity in disc evolution remain contentious. Dolidze 25 is an excellent laboratory to study star formation at lower metallicities. The disc fraction of Dolidze 25 is smaller than what would be predicted based on its age alone \citep{Guarcello2021}. Given the low stellar density of the cluster and low O-type stellar population, the rapid disk dispersal is likely governed by the low metallicity of the complex rather than external photoevaporation or dynamical interactions, which are proposed as the major external factors influencing the disk evolution \citep[e.g.][]{Winter2018}. Hence it is crucial to study how the low metallicity influences disk evolution in this region using state of art instruments and techniques. \par

In this paper, we discuss the preliminary results of the spectroscopic analysis of the disk population in the central region of Dolidze 25, a low-metallicity young cluster, using medium-resolution spectra extracted from VLT-MUSE data. We obtained the flux-calibrated spectra of 280 stars in the field corrected for the average value of radial velocity (71 km/s) of the cluster from literature \citep{Wu2009}. The accretors are selected based on three criteria, the equivalent width of H$\alpha$ emission $>$ 3 $\AA$, velocity broadened H$\alpha$ profiles  with 10\% width $>$ 270 km/s, as well as the existence of emission lines such as [OI] and $H\beta$. Based on this we identify 14 accreting sources within the field. Of them, six sources are confirmed disk sources based on their IR excess. We also confirm that the 14 accreting sources in Do 25 belong to the pre-main sequence class based on their locations on the optical and IR color-magnitude diagrams. \par

Our work aims to achieve long-term goals, which involve estimating the physical characteristics of sources in Dolidze 25, such as age, mass, distance, and luminosity, as well as conducting membership analysis. Furthermore, we intend to assess the accretion rate of identified accretors and compare them with solar metallicity spectral templates of nearby regions, such as Orion and NGC 2264, to determine how physical parameters vary with metallicity. To attain these goals, we are conducting an in-depth investigation that employs low-metallicity models to establish a correlation between accretion and stellar parameters. The findings of our research will be reported in a subsequent publication.

\bibliography{main}

\begin{thebibliography}{}
\expandafter\ifx\csname natexlab\endcsname\relax\def\natexlab#1{#1}\fi

\bibitem[{{Andrews}(2020)}]{Andrews2020}
{Andrews}, S.~M. 2020, ARAA, 58, 483

\bibitem[{{Barentsen} {$et~al$.}(2014){Barentsen}, {Farnhill}, {Drew},
  {Gonz{\'a}lez-Solares}, {Greimel}, {Irwin}, {Miszalski}, {Ruhland}, {Groot},
  {Mampaso}, {Sale}, {Henden}, {Aungwerojwit}, {Barlow}, {Carter}, {Corradi},
  {Drake}, {Eisl{\"o}ffel}, {Fabregat}, {G{\"a}nsicke}, {Gentile Fusillo},
  {Greiss}, {Hales}, {Hodgkin}, {Huckvale}, {Irwin}, {King}, {Knigge},
  {Kupfer}, {Lagadec}, {Lennon}, {Lewis}, {Mohr-Smith}, {Morris}, {Naylor},
  {Parker}, {Phillipps}, {Pyrzas}, {Raddi}, {Roelofs}, {Rodr{\'\i}guez-Gil},
  {Sabin}, {Scaringi}, {Steeghs}, {Suso}, {Tata}, {Unruh}, {van Roestel},
  {Viironen}, {Vink}, {Walton}, {Wright}, \& {Zijlstra}}]{IPHAS2014}
{Barentsen}, G., {Farnhill}, H.~J., {Drew}, J.~E., {$et~al$.} 2014, MNRAS, 444,
  3230

\bibitem[{{Bressan} {$et~al$.}(2012){Bressan}, {Marigo}, {Girardi},
  {Salasnich}, {Dal Cero}, {Rubele}, \& {Nanni}}]{Parsec2012}
{Bressan}, A., {Marigo}, P., {Girardi}, L., {$et~al$.} 2012, MNRAS, 427, 127

\bibitem[{{Cabrit} {$et~al$.}(1990){Cabrit}, {Edwards}, {Strom}, \&
  {Strom}}]{1990Cabrit}
{Cabrit}, S., {Edwards}, S., {Strom}, S.~E., \& {Strom}, K.~M. 1990, ApJ, 354,
  687

\bibitem[{{Chambers} {$et~al$.}(2016){Chambers}, {Magnier}, {Metcalfe},
  {Flewelling}, {Huber}, {Waters}, {Denneau}, {Draper}, {Farrow}, {Finkbeiner},
  {Holmberg}, {Koppenhoefer}, {Price}, {Rest}, {Saglia}, {Schlafly}, {Smartt},
  {Sweeney}, {Wainscoat}, {Burgett}, {Chastel}, {Grav}, {Heasley}, {Hodapp},
  {Jedicke}, {Kaiser}, {Kudritzki}, {Luppino}, {Lupton}, {Monet}, {Morgan},
  {Onaka}, {Shiao}, {Stubbs}, {Tonry}, {White}, {Ba{\~n}ados}, {Bell},
  {Bender}, {Bernard}, {Boegner}, {Boffi}, {Botticella}, {Calamida},
  {Casertano}, {Chen}, {Chen}, {Cole}, {Deacon}, {Frenk}, {Fitzsimmons},
  {Gezari}, {Gibbs}, {Goessl}, {Goggia}, {Gourgue}, {Goldman}, {Grant},
  {Grebel}, {Hambly}, {Hasinger}, {Heavens}, {Heckman}, {Henderson}, {Henning},
  {Holman}, {Hopp}, {Ip}, {Isani}, {Jackson}, {Keyes}, {Koekemoer}, {Kotak},
  {Le}, {Liska}, {Long}, {Lucey}, {Liu}, {Martin}, {Masci}, {McLean}, {Mindel},
  {Misra}, {Morganson}, {Murphy}, {Obaika}, {Narayan}, {Nieto-Santisteban},
  {Norberg}, {Peacock}, {Pier}, {Postman}, {Primak}, {Rae}, {Rai}, {Riess},
  {Riffeser}, {Rix}, {R{\"o}ser}, {Russel}, {Rutz}, {Schilbach}, {Schultz},
  {Scolnic}, {Strolger}, {Szalay}, {Seitz}, {Small}, {Smith}, {Soderblom},
  {Taylor}, {Thomson}, {Taylor}, {Thakar}, {Thiel}, {Thilker}, {Unger},
  {Urata}, {Valenti}, {Wagner}, {Walder}, {Walter}, {Watters}, {Werner},
  {Wood-Vasey}, \& {Wyse}}]{Panstarrs2016}
{Chambers}, K.~C., {Magnier}, E.~A., {Metcalfe}, N., {$et~al$.} 2016, arXiv
  e-prints, arXiv:1612.05560

\bibitem[{{De Marchi} {$et~al$.}(2013){De Marchi}, {Beccari}, \&
  {Panagia}}]{2013Demarchi}
{De Marchi}, G., {Beccari}, G., \& {Panagia}, N. 2013, ApJ, 775, 68

\bibitem[{{Fang} {$et~al$.}(2009){Fang}, {van Boekel}, {Wang}, {Carmona},
  {Sicilia-Aguilar}, \& {Henning}}]{Fang2009}
{Fang}, M., {van Boekel}, R., {Wang}, W., {$et~al$.} 2009, A\&A, 504, 461

\bibitem[{{Fang} {$et~al$.}(2012){Fang}, {van Boekel}, {King}, {Henning},
  {Bouwman}, {Doi}, {Okamoto}, {Roccatagliata}, \&
  {Sicilia-Aguilar}}]{Fang2012}
{Fang}, M., {van Boekel}, R., {King}, R.~R., {$et~al$.} 2012, A\&A, 539, A119

\bibitem[{{Fedele} {$et~al$.}(2010){Fedele}, {van den Ancker}, {Henning},
  {Jayawardhana}, \& {Oliveira}}]{Fedele2010}
{Fedele}, D., {van den Ancker}, M.~E., {Henning}, T., {Jayawardhana}, R., \&
  {Oliveira}, J.~M. 2010, A\&A, 510, A72

\bibitem[{{Fischer} {$et~al$.}(2022){Fischer}, {Hillenbrand}, {Herczeg},
  {Johnstone}, {K{\'o}sp{\'a}l}, \& {Dunham}}]{Fischer2022}
{Fischer}, W.~J., {Hillenbrand}, L.~A., {Herczeg}, G.~J., {$et~al$.} 2022,
  arXiv e-prints, arXiv:2203.11257

\bibitem[{{Fitzsimmons} {$et~al$.}(1992){Fitzsimmons}, {Dufton}, \&
  {Rolleston}}]{Fitzsimmons1992}
{Fitzsimmons}, A., {Dufton}, P.~L., \& {Rolleston}, W.~R.~J. 1992, MNRAS, 259,
  489

\bibitem[{{Ginsburg} {$et~al$.}(2022){Ginsburg}, {Sokolov}, {de Val-Borro},
  {Rosolowsky}, {Pineda}, {Sip{\H{o}}cz}, \& {Henshaw}}]{pyspeckit}
{Ginsburg}, A., {Sokolov}, V., {de Val-Borro}, M., {$et~al$.} 2022, AJ, 163,
  291

\bibitem[{{Guarcello} {$et~al$.}(2021){Guarcello}, {Biazzo}, {Drake}, {Micela},
  {Prisinzano}, {Sciortino}, {Damiani}, {Flaccomio}, {Neiner}, \&
  {Wright}}]{Guarcello2021}
{Guarcello}, M.~G., {Biazzo}, K., {Drake}, J.~J., {$et~al$.} 2021, A\&A, 650,
  A157

\bibitem[{{Hartigan} {$et~al$.}(1995){Hartigan}, {Edwards}, \&
  {Ghandour}}]{Hartigan1995}
{Hartigan}, P., {Edwards}, S., \& {Ghandour}, L. 1995, ApJ, 452, 736

\bibitem[{{Hartmann} {$et~al$.}(1998){Hartmann}, {Calvet}, {Gullbring}, \&
  {D'Alessio}}]{1998Hartmann}
{Hartmann}, L., {Calvet}, N., {Gullbring}, E., \& {D'Alessio}, P. 1998, ApJ,
  495, 385

\bibitem[{{Hartmann} {$et~al$.}(2016){Hartmann}, {Herczeg}, \&
  {Calvet}}]{Hartmann2016}
{Hartmann}, L., {Herczeg}, G., \& {Calvet}, N. 2016, ARAA, 54, 135

\bibitem[{{Kalari} \& {Vink}(2015)}]{Kalari2015}
{Kalari}, V.~M., \& {Vink}, J.~S. 2015, ApJ, 800, 113

\bibitem[{{Kurosawa} \& {Proga}(2008)}]{Kurosawa2008}
{Kurosawa}, R., \& {Proga}, D. 2008, ApJ, 674, 97

\bibitem[{{Lawrence} {$et~al$.}(2007){Lawrence}, {Warren}, {Almaini}, {Edge},
  {Hambly}, {Jameson}, {Lucas}, {Casali}, {Adamson}, {Dye}, {Emerson},
  {Foucaud}, {Hewett}, {Hirst}, {Hodgkin}, {Irwin}, {Lodieu}, {McMahon},
  {Simpson}, {Smail}, {Mortlock}, \& {Folger}}]{UKIDSS2007}
{Lawrence}, A., {Warren}, S.~J., {Almaini}, O., {$et~al$.} 2007, MNRAS, 379,
  1599

\bibitem[{{Lennon} {$et~al$.}(1990){Lennon}, {Dufton}, {Fitzsimmons}, {Gehren},
  \& {Nissen}}]{Lennon1990}
{Lennon}, D.~J., {Dufton}, P.~L., {Fitzsimmons}, A., {Gehren}, T., \& {Nissen},
  P.~E. 1990, A\&A, 240, 349

\bibitem[{{Manara} {$et~al$.}(2022){Manara}, {Ansdell}, {Rosotti}, {Hughes},
  {Armitage}, {Lodato}, \& {Williams}}]{Manara2022}
{Manara}, C.~F., {Ansdell}, M., {Rosotti}, G.~P., {$et~al$.} 2022, arXiv
  e-prints, arXiv:2203.09930

\bibitem[{{Manoj} {$et~al$.}(2006){Manoj}, {Bhatt}, {Maheswar}, \&
  {Muneer}}]{2006Manoj}
{Manoj}, P., {Bhatt}, H.~C., {Maheswar}, G., \& {Muneer}, S. 2006, ApJ, 653,
  657

\bibitem[{{McLeod} {$et~al$.}(2020){McLeod}, {Kruijssen}, {Weisz}, {Zeidler},
  {Schruba}, {Dalcanton}, {Longmore}, {Chevance}, {Faesi}, \&
  {Byler}}]{Anna2020}
{McLeod}, A.~F., {Kruijssen}, J.~M.~D., {Weisz}, D.~R., {$et~al$.} 2020, ApJ,
  891, 25

\bibitem[{{Muzerolle} {$et~al$.}(2004){Muzerolle}, {D'Alessio}, {Calvet}, \&
  {Hartmann}}]{2004Muzerolle}
{Muzerolle}, J., {D'Alessio}, P., {Calvet}, N., \& {Hartmann}, L. 2004, ApJ,
  617, 406

\bibitem[{{Muzerolle} {$et~al$.}(2003){Muzerolle}, {Hillenbrand}, {Calvet},
  {Brice{\~n}o}, \& {Hartmann}}]{Muzerolle2003}
{Muzerolle}, J., {Hillenbrand}, L., {Calvet}, N., {Brice{\~n}o}, C., \&
  {Hartmann}, L. 2003, ApJ, 592, 266

\bibitem[{{Negueruela} {$et~al$.}(2015){Negueruela}, {Sim{\'o}n-D{\'\i}az},
  {Lorenzo}, {Castro}, \& {Herrero}}]{Negueruela2015}
{Negueruela}, I., {Sim{\'o}n-D{\'\i}az}, S., {Lorenzo}, J., {Castro}, N., \&
  {Herrero}, A. 2015, A\&A, 584, A77

\bibitem[{{N{\'u}{\~n}ez} {$et~al$.}(2022){N{\'u}{\~n}ez}, {Douglas}, {Alam},
  \& {DeLaurentiis}}]{PHEW2022}
{N{\'u}{\~n}ez}, A., {Douglas}, S.~T., {Alam}, M., \& {DeLaurentiis}, S. 2022,
  {PHEW: PytHon Equivalent Widths}, Zenodo, doi:10.5281/zenodo.6422571

\bibitem[{{Puga} {$et~al$.}(2007){Puga}, {Neiner}, {Hony}, {Lenorzer},
  {Hubert}, \& {Waters}}]{Puga2007}
{Puga}, E., {Neiner}, C., {Hony}, S., {$et~al$.} 2007, in Triggered Star
  Formation in a Turbulent ISM, ed. B.~G. {Elmegreen} \& J.~{Palous}, Vol. 237,
  465--465

\bibitem[{{Spezzi} {$et~al$.}(2012){Spezzi}, {De Marchi}, {Panagia},
  {Sicilia-Aguilar}, \& {Ercolano}}]{2012Spezzi}
{Spezzi}, L., {De Marchi}, G., {Panagia}, N., {Sicilia-Aguilar}, A., \&
  {Ercolano}, B. 2012, MNRAS, 421, 78

\bibitem[{{Weilbacher} {$et~al$.}(2014){Weilbacher}, {Streicher}, {Urrutia},
  {P{\'e}contal-Rousset}, {Jarno}, \& {Bacon}}]{Weilbacher2014}
{Weilbacher}, P.~M., {Streicher}, O., {Urrutia}, T., {$et~al$.} 2014, in
  Astronomical Society of the Pacific Conference Series, Vol. 485, Astronomical
  Data Analysis Software and Systems XXIII, ed. N.~{Manset} \& P.~{Forshay},
  451

\bibitem[{{White} \& {Basri}(2003)}]{White2003}
{White}, R.~J., \& {Basri}, G. 2003, ApJ, 582, 1109

\bibitem[{{Winter} {$et~al$.}(2018){Winter}, {Clarke}, {Rosotti}, {Ih},
  {Facchini}, \& {Haworth}}]{Winter2018}
{Winter}, A.~J., {Clarke}, C.~J., {Rosotti}, G., {$et~al$.} 2018, MNRAS, 478,
  2700

\bibitem[{{Wu} {$et~al$.}(2009){Wu}, {Zhou}, {Ma}, \& {Du}}]{Wu2009}
{Wu}, Z.-Y., {Zhou}, X., {Ma}, J., \& {Du}, C.-H. 2009, MNRAS, 399, 2146

\bibitem[{{Yasui} {$et~al$.}(2016{\natexlab{a}}){Yasui}, {Kobayashi}, {Saito},
  \& {Izumi}}]{2016bYasui}
{Yasui}, C., {Kobayashi}, N., {Saito}, M., \& {Izumi}, N. 2016{\natexlab{a}},
  AJ, 151, 115

\bibitem[{{Yasui} {$et~al$.}(2021){Yasui}, {Kobayashi}, {Saito}, {Izumi}, \&
  {Skidmore}}]{2021Yasui}
{Yasui}, C., {Kobayashi}, N., {Saito}, M., {Izumi}, N., \& {Skidmore}, W. 2021,
  AJ, 161, 139

\bibitem[{{Yasui} {$et~al$.}(2016{\natexlab{b}}){Yasui}, {Kobayashi},
  {Tokunaga}, {Saito}, \& {Izumi}}]{2016aYasui}
{Yasui}, C., {Kobayashi}, N., {Tokunaga}, A.~T., {Saito}, M., \& {Izumi}, N.
  2016{\natexlab{b}}, AJ, 151, 50

\bibitem[{{Yasui} {$et~al$.}(2010){Yasui}, {Kobayashi}, {Tokunaga}, {Saito}, \&
  {Tokoku}}]{Yasui2009}
{Yasui}, C., {Kobayashi}, N., {Tokunaga}, A.~T., {Saito}, M., \& {Tokoku}, C.
  2010, ApJl, 723, L113

\bibitem[{{Zeidler}(2019)}]{Peter2019}
{Zeidler}, P. 2019, {MUSEpack}, Zenodo, doi:10.5281/zenodo.3433996

\end{thebibliography}
\end{document}